\documentclass[12pt]{article}
\usepackage{verbatim,color,amssymb,epsfig}
\usepackage[usenames,dvipsnames]{xcolor}
\usepackage{fancyhdr}
\usepackage[authoryear, sort]{natbib}
\usepackage{amsmath}
\usepackage[hidelinks]{hyperref}
\usepackage[labelfont=bf]{caption}
\usepackage{floatrow}
\usepackage{setspace}
\usepackage{cleveref}
\usepackage{soul}

\crefname{Rem}{Remark}{Temarks}
\crefname{Thm}{Theorem}{Theorems}
\crefname{section}{Section}{Sections}
\crefname{table}{Table}{Tables}

\renewcommand\footnoterule{\kern-3pt \hrule \textwidth 2in \kern 2.6pt}

\newtheorem{Rem}{\underline{\bf Remark}}
\newtheorem{Thm}{\underline{\bf Theorem}}

\def\L{{\cal L}}

\def\b1e{{\mathbf e}}

\def\bbeta{\mbox{\boldmath $\beta$}}

\def\wh{\widehat}

\def\cov{\hbox{\rm cov}}
\def\diag{\hbox{diag}}
\def\log{\hbox{log}}

\def\Normal{\hbox{Normal}}
\def\pr{\hbox{pr}}

\def\sumd{\hbox{$\sum_{d=0}^{1}$}}
\def\sumi{\hbox{$\sum_{i=1}^n$}}

\def\sumjn{\hbox{$\sum_{j=1}^n$}}

\def\sumr{\hbox{$\sum_{r=0}^{1}$}}

\def\be{\begin{eqnarray}}
\def\ee{\end{eqnarray}}
\def\bq{\begin{equation}}
\def\eq{\end{equation}}
\def\bse{\begin{eqnarray*}}
\def\ese{\end{eqnarray*}}

\def\Supp{\emph{Supplementary Material}}

\def\trans{^{\rm T}}

\def\authorfootnote#1{{\let\thefootnote\relax\footnotetext{#1}}}
\def\boxit#1{\vbox{\hrule\hbox{\vrule\kern6pt \vbox{\kern6pt \textcolor{blue}{#1}\kern6pt}\kern6pt\vrule}\hrule}}

\begin{document}

\thispagestyle{empty}
	\baselineskip=28pt
	\thispagestyle{empty}
	\baselineskip=28pt
	{\LARGE{\bf \begin{center}
				Improved Semiparametric Analysis of Polygenic Gene-Environment Interactions in Case-Control Studies
			\end{center}  }}
			
			\baselineskip=12pt
			
			\vskip 2mm
			\begin{center}
				Tianying Wang	\footnote{Corresponding author at: Weiqinglou Rm 212-A, Center for Statistical Science, Tsinghua University, Beijing 100084, China.
E-mail address: tianyingw@tsinghua.edu.cn (T. Wang). }\\
				Center for Statistical Science, Tsinghua University,\\
				Department of Industrial Engineering, Tsinghua University\\
				Beijing, China \\
				tianyingw@tsinghua.edu.cn \\
				\hskip 5mm\\
				Alex Asher\\
				StataCorp LLC, \\
				4905 Lakeway Dr, College Station, TX 77845, U.S.A\\ 
				aasher@stata.com\\
			\end{center}

			\begin{center}
				{\Large{\bf Abstract}}
			\end{center}
			\baselineskip=12pt
			
			Standard logistic regression analysis of case-control data has low power to detect gene-environment interactions, but until recently it was the only method that could be used on complex polygenic data for which parametric distributional models are not feasible.  Under the assumption of gene-environment independence in the underlying population, \citeauthor{Stalder2017} (2017, \emph{Biometrika}, {\bf 104}, 801-812) developed a retrospective method that treats both genetic and environmental variables nonparametrically. However, the mathematical symmetry of genetic and environmental variables is overlooked.  We propose an improvement to the method of \citeauthor{Stalder2017} that increases the efficiency of the estimates with no additional assumptions and modest computational cost.  This improvement is achieved by treating the genetic and environmental variables symmetrically to generate two sets of parameter estimates that are combined to generate a more efficient estimate.  We employ a semiparametric framework to develop the asymptotic theory of the estimator, {showed its asymptotic efficiency gain,} and evaluate its performance via simulation studies.  The method is illustrated using data from a case-control study of breast cancer.

			\baselineskip=12pt
			\par\vfill\noindent
			\underline{\bf Some Key Words}:
			Gene-environment interaction;
			Case-control study;
			Genetic epidemiology;
			Semiparametric estimation;
			Biased samples.

			\par\medskip\noindent
			\underline{\bf Short title}: Semiparametric Analysis of Gene-Environment Interactions

			\clearpage\pagebreak\newpage
			\pagenumbering{arabic}
			\newlength{\gnat}
			\setlength{\gnat}{22pt}
			\baselineskip=\gnat

\section{Introduction}\label{sec:Intro}

Genetic epidemiologists have identified both genetic and environmental factors that influence the incidence of complex diseases such as cancers, heart diseases, depression, and diabetes \citep{Nickels2013interaction, Rudolph2015interaction, mullins2016depression, Gustavsson2016FTO, Krischer2017genetic}.  As new studies identify additional genetic variants associated with a disease, attention turns to exploring the interaction between genetic susceptibility and environmental risk factors.

Researchers studying gene-environment interactions often adopt a case-control study design, wherein diseased cases and healthy control subjects are identified and their covariate information is collected retrospectively.  When the disease is rare, sampling cases and controls separately provides substantial cost and time savings over a prospective cohort study, but it makes statistical inference more complicated.

\citet{PrenticePyke1979} demonstrated that standard prospective logistic regression of case-control data, which ignores the retrospective sampling scheme, nevertheless yields consistent estimates of all parameters except the logistic intercept.  Logistic regression is equivalent to maximum likelihood estimation under a model that places no assumptions on the joint distribution of the genetic and environmental variables, and it achieves the variance lower bound under this model \citep{Breslow2000}. However, logistic regression of case-control data lacks power to detect gene-environment interaction effects.

To improve estimation efficiency, studies of gene-environment interactions often take advantage of the relatively mild assumption that the genetic and environmental variables are independently distributed in the source population.  This assumption is easy to test, is frequently valid, and enables the use of specialized methods for the analysis of case-control data.  \Citet{Piegorsch1994} proposed a case-only approach that efficiently estimates multiplicative interactions (but not main effects) under the assumptions of gene-environment independence and rare disease.  \Citet{ChatterjeeCarroll2005} exploited the gene-environment independence assumption to develop a semiparametric retrospective profile likelihood framework that treats environmental variables nonparametrically but assumes that the genetic variables have a known, discrete distribution.  Further developments have yielded additional retrospective methods based on parametric modeling of the distribution of genetic variables given the environmental variables, see for example \cite{Lobach2008, Ma2010, han2012likelihood, liang2019semiparametric}.

Genome-wide association studies have shown that genetic predisposition to a single disease tends to be highly polygenic, with many genetic variants influencing disease risk \citep{chatterjee2016developing, Fuchsberger2016}.  To provide a more complete picture of genetic risk and gene-environment interactions, it is often advantageous to include multiple genetic loci in the disease model \citep{chatterjee2006powerful, jiao2013sberia, lin2015test}.  In the interest of parsimony, many studies have focused on developing polygenic risk scores through a weighted combination of all known genetic variants associated with a disease \citep{ChatterjeeWheeler2013polygenic, Dudbridge2013}.  Handling multiple genetic variants, polygenic risk scores, or a combination of both is straightforward with prospective logistic regression, but can be unwieldy or even impossible when using retrospective methods that exploit gene-environment independence to gain efficiency but require a parametric model for the distribution of the genetic component.

The method of \citet{Stalder2017} extends the Chatterjee-Carroll retrospective profile likelihood framework by treating both the genetic and environmental variables nonparametrically, requiring only the assumption of gene-environment independence in the source population.  This assumption of independence can be weakened if a discrete stratification variable is found such that genes and environment are independent within strata of the source population. However, \citet{Stalder2017} overlooked the mathematical symmetry of the genetic and environmental variables in the retrospective likelihood, resulting in a sub-optimal efficiency gain.

With this valuable discovery, here we propose an improvement to the method of \citet{Stalder2017} that substantially increases the efficiency of the estimates with no additional assumptions and modest computational cost.  This development relies on the observation that the method of \citeauthor{Stalder2017} removes dependence on the distribution of the genetic and environmental variables in two different fashions; by treating the genetic and environmental variables symmetrically we generate two sets of parameter estimates that are combined to generate a more efficient estimate.  We employ a semiparametric framework to develop the asymptotic theory of the estimator.  The properties of the new method are illustrated through simulations in \cref{sec:Simulations}, and an example in \cref{sec:DataAnalysis}.

\section{Methodology and Theory}\label{sec:MethodTheory}

\subsection{Background}\label{sec:Background}

We adopt notation similar to that of \citeauthor{Stalder2017}, with disease status, genetic information, and environmental risk factors denoted by $D$, $G$, and $X$, respectively.  Both $G$ and $X$ are potentially multivariate and can contain both discrete and continuous components.  For a given case-control study, $n_1$ is the number of cases ($D=1$) and $n_0$ is the number of controls ($D=0$), while $\pi_1 = \pr(D=1)$ is the disease rate in the source population and $\pi_0 = 1 - \pi_1$. We maintain the assumption in \citeauthor{Stalder2017} that $\pi_1$ is either known or can be estimated well.

The assumption of gene-environment independence in the source population can be written as $f_{GX}(g, x) = f_{G}(g) \times f_{X}(x)$, where $f_{GX}(\cdot, \cdot)$ is the joint density or mass of $G$ and $X$ in the underlying population, and $f_{G}(\cdot)$ and $f_{X}(\cdot)$ are the marginal density or mass functions of $G$ and $X$, respectively, in the underlying population. We assume $f_X(x)$ and $f_G(g)$ are completely unspecified.

Given the genetic and environmental covariates, we assume the risk of disease in the underlying population follows the model $\pr(D=1 \mid G,X) = H\{\alpha_0 + m(G,X,\bbeta)\},$ where $H(x)=\{ 1 + \exp(-x)\}^{-1}$ is the logistic distribution function and $m(G,X,\bbeta)$ is a function that describes the joint effect of $G$ and $X$ and is known up to the unspecified parameters of interest $\bbeta$.  

Given the subject's disease status, the retrospective likelihood is the probability of observing the genetic and environmental variables.  Under gene-environment independence in the source population,
the retrospective likelihood is 
\begin{equation*}
  \frac{f_G(g) f_X(x) \exp[d\{\alpha_0 + m(g,x,\beta)\}]/ [1 + \exp\{\alpha_0 + m(g,x,\beta)\}]}
       {\int f_G(u) f_X(v) \exp[d\{\alpha_0 + m(u,v,\beta)\}]/ [1 + \exp\{\alpha_0 + m(u,v,\beta)\}] du dv}.
  \label{eq:RetroLik}
\end{equation*}
The logistic intercept $\alpha_0$, typically of little scientific interest, is not consistently estimated using prospective logistic regression, which instead converges to $\kappa = \alpha_0 + \log(n_1/n_0) - \log(\pi_1/\pi_0)$ \citep{PrenticePyke1979}.  For convenience, we parameterize everything in terms of $\kappa$, and we define $\Omega = (\kappa,\bbeta\trans)\trans$. 
\citet{ChatterjeeCarroll2005} profiled out $f_X(\cdot)$ to create a semiparametric profile likelihood
\begin{equation}
  L_X(D,G,X,\Omega,f_{G}) = f_G(G) \frac{S(D,G,X,\Omega)}{R_X(X,\Omega)}, \label{eq:ProfLik_X}
\end{equation}
where
\begin{align}
  S(d,g,x,\Omega) &= \frac{\exp[d\{\kappa + m(g,x,\beta)\}]}{1 + \exp\{\kappa - \log(n_1/n_0) + \log(\pi_1/\pi_0) + m(g,x,\beta)\}}; \nonumber \\
  R_X(x,\Omega) &= \sumr \int f_{G}(v) S(r,v,x,\Omega) dv. \label{eq:R_X}
\end{align}

The key insight of \citet{Stalder2017} was to develop an unbiased estimator of $R_X(x,\Omega)$ that treats $f_G(\cdot)$ nonparametrically, defined as
\begin{equation}
  \wh{R}_X(x,\Omega) = \sumjn \sumr \sumd (\pi_d / n_d)  I(D_j=d)  S(r,G_j,x,\Omega). \label{eq:Rhat_X}
\end{equation}
The leading term $f_G(G)$ in \cref{eq:ProfLik_X} is constant with respect to $\Omega$, and can be ignored for the purpose of estimation.  Replacing $R_X(x,\Omega)$ with $\wh{R}_X(x,\Omega)$ and taking the logarithm yields the estimated profile loglikelihood of $\Omega$ given the data as
\begin{equation}
  \wh{\L}_X(\Omega) = \sumi \log\{S(D_i,G_i,X_i,\Omega)\} - \sumi \log\{\wh{R}_{X}(X_i,\Omega)\}. \label{eq:Lik_X} 
\end{equation}

Define $S_{\Omega}(d,g,x,\Omega) = \partial S(d,g,x,\Omega) / \partial \Omega$ and $\wh{R}_{X \Omega}(x,\Omega) = \partial \wh{R}_{X}(x, \Omega) / \partial \Omega$.  The profile likelihood score function, ${\cal S}_{X}(\Omega)$, is unknown but can be estimated consistently by
\begin{equation}
  \wh{{\cal S}}_{X}(\Omega) = n^{-1/2}\sum_{i=1}^{n} \left\{\frac{S_{\Omega}(D_i,G_i,X_i,\Omega)}{S(D_i,G_i,X_i,\Omega)} -\frac{\wh{R}_{X\Omega}(X_i,\Omega)}{\wh{R}_{X}(X_i,\Omega)}\right\}. \label{eq:ScoreHat_X}
\end{equation}
By solving $\wh{{\cal S}}_{X}(\Omega)=0$, we obtain a consistent estimate of $\Omega$, which we will denote as $\wh{\Omega}_X$ and which is called the SPMLE by \citet{Stalder2017}.

\subsection{Symmetric Combination Estimator}\label{sec:Symm}

The above equations are equivalent to those found in \citet{Stalder2017} with the addition of the subscript $X$ in \crefrange{eq:ProfLik_X}{eq:ScoreHat_X} to emphasize that the density of $X$ has been profiled out, leaving the density of $G$ to be treated nonparametrically.  Because our only assumption about $G$ and $X$ is their independence in the source population, we could just as well have interchanged them and profiled out the distribution of $G$.  The notation in this symmetric case is analogous to the above, but with subscript $G$ instead of $X$.  
It follows that the analogous estimated score function
\begin{equation*}
  \wh{{\cal S}}_{G}(\Omega) = n^{-1/2}\sum_{i=1}^{n}\left\{ \frac{S_{\Omega}(D_i,G_i,X_i,\Omega)}{S(D_i,G_i,X_i,\Omega)} -  \frac{\wh{R}_{G \Omega}(G_i,\Omega)}{\wh{R}_{G}(G_i,\Omega)}\right\} \label{eq:ScoreHat_G}
\end{equation*}
can be used to obtain $\wh{\Omega}_G$, another consistent estimate of $\Omega$.

The optimal combination of symmetric estimators $\wh{\Omega}_{X}$ and $\wh{\Omega}_{G}$ follows the principle of generalized least squares.  Suppose the dimension of $\Omega$ is $p$.  Let $I_p$ be the $p\times p$ identity matrix and define ${\cal X} = (I_p,I_p)\trans$.  Define ${\cal Y} = (\wh{\Omega}_X\trans,\wh{\Omega}_G\trans)\trans$ and $\Lambda_{\rm all} = \cov({\cal Y})$.  \Cref{thm:AsympNormSymm}, in \cref{sec:AsympTheory}, shows ${\cal Y} \to \Normal({\cal X}\Omega, \Lambda_{\rm all})$.

Treating this as a generalized least squares problem, we can rewrite it as ${\cal Y} = {\cal X}\Omega+\epsilon,$ where $\epsilon \sim \Normal (0, \Lambda_{\rm all})$.  The Symmetric Combination Estimator is the solution to the linear model, namely
\begin{equation}
  \wh{\Omega}_{\rm Symm} = ({\cal X}\trans \Lambda^{-1}_{\rm all}{\cal X})^{-1}{\cal X}\trans \Lambda^{-1}_{\rm all}{\cal Y}. \label{eq:OmegaHatSymm}
\end{equation}

An alternative method of combining the two estimates is to average the two estimated profile likelihoods into a single composite likelihood.  The resulting Composite Likelihood Estimator yields minimal efficiency gains over the SPMLE from \citeauthor{Stalder2017}, and is presented in \cref{sec:CompositeLik} of the \Supp{}.

\subsection{Asymptotic Theory}\label{sec:AsympTheory}

In this subsection we first demonstrate that the joint distribution of $\wh\Omega_X$ and $\wh\Omega_G$ is asymptotically normal.  We then show the asymptotic results for the Symmetric Combination Estimator.  In practice, asymptotic standard errors for the Symmetric Combination Estimator proved unreliable due to slow convergence, so bootstrap standard errors are used instead.

To state the asymptotic results, let $Z_i = (D_i,G_i,X_i)$ then define
\bse
  \Gamma_{X} &=& \sum_{d=0}^{1} \frac{n_d}{n} E\left[ \frac{\partial}{\partial \Omega\trans} \left\{ \frac{S_{\Omega}(Z, \Omega)}{S(Z, \Omega)} - \frac{R_{X \Omega}(X,\Omega)}{R_{X}(X,\Omega)}  \right\} \bigg| D=d \right]; \\
  \Gamma_{G} &=& \sum_{d=0}^{1} \frac{n_d}{n} E\left[ \frac{\partial}{\partial \Omega\trans} \left\{ \frac{S_{\Omega}(Z, \Omega)}{S(Z, \Omega)} - \frac{R_{G \Omega}(X,\Omega)}{R_{G}(X,\Omega)}  \right\} \bigg| D=d \right]; \\
  \zeta_{X}(Z_i,\Omega) &=&\frac{S_{\Omega}(Z_i,\Omega)}{S(Z_i,\Omega)} - \frac{R_{X \Omega}(X_i,\Omega)}{R_{X}(X_i,\Omega)}\\
                         && -\sum_{d=0}^1\sum_{r=0}^1 \frac{n_d\pi_{d_i}}{n_{d_i}} E\left\{\frac{S_\Omega(r,g,X,\Omega)}{R_{X}(X, \Omega)} -\frac{R_{X \Omega}(X, \Omega)S(r,g,X,\Omega)}{R^2_{X}(X, \Omega)} \bigg| D=d\right\}_{g=G_i};\\
  \zeta_{G}(Z_i,\Omega) &=& \frac{S_{\Omega}(Z_i,\Omega)}{S(Z_i,\Omega)} - \frac{R_{G \Omega}(G_i,\Omega)}{R_{G}(G_i,\Omega)}\\
                         && -\sum_{d=0}^1\sum_{r=0}^1 \frac{n_d\pi_{d_i}}{n_{d_i}} E\left\{\frac{S_\Omega(r,G,x,\Omega)}{R_{G}(G, \Omega)} -\frac{R_{G \Omega}(G, \Omega)S(r,G,x,\Omega)}{R^2_{G}(G, \Omega)} \bigg| D=d\right\}_{x=X_i};\\
  \zeta_{X*}(Z_i,\Omega) &=& \zeta_{X}(Z_i,\Omega) - E\{\zeta_{X}(Z,\Omega) \vert D=D_i\};\\
  \zeta_{G*}(Z_i,\Omega) &=& \zeta_{G}(Z_i,\Omega) - E\{\zeta_{G}(Z,\Omega) \vert D=D_i\}.
\ese

By profiling $X$ and $G$ out separately, we have the following two equations
\begin{align}
  n^{1/2}(\wh\Omega_X-\Omega)&=-\Gamma_X^{-1}n^{-1/2}\sumi \zeta_{X*}(Z_i,\Omega)+o_p(1);\label{eq:asymp_X}\\
  n^{1/2}(\wh\Omega_G-\Omega)&=-\Gamma_G^{-1}n^{-1/2}\sumi \zeta_{G*}(Z_i,\Omega)+o_p(1).\label{eq:asymp_G}
\end{align}
\Cref{eq:asymp_X} is proved in Theorem 1 of \citet{Stalder2017}, and the proof of the symmetric case in \cref{eq:asymp_G} is analogous.

To demonstrate the asymptotic properties of the Symmetric Combination Estimator, denote the block-diagonal matrix $\Gamma_{\rm all}^{-1}=\diag(\Gamma_{X}^{-1}, \Gamma_{G}^{-1})$.
\begin{Thm}\label{thm:AsympNormSymm}{
  Suppose that $0 < \displaystyle\lim_{n\to\infty} n_d/n<1$, and $\pi_1$ is known. Then
  \begin{equation*}
    n^{1/2}({\cal Y}-{\cal X}\Omega) = -\Gamma_{\rm all}^{-1} \;n^{-1/2} \; \sum_{i=1}^{n} \left\{\begin{matrix} \zeta_{X*}(Z_i,\Omega) \\ \zeta_{G*}(Z_i,\Omega) \end{matrix}\right\} + o_p(1).
  \end{equation*}
  The $Z_i$ are independent and $E\{\zeta_{X*}(Z_i,\Omega)\vert D_i\} = E\{\zeta_{G*}(Z_i,\Omega)\vert D_i\} = 0$, so as $n \to \infty,$
  \begin{equation}
    n^{1/2}({\cal Y}-{\cal X}\Omega) \to \Normal(0,\Lambda_{\rm all}) \label{eq:AsympNormSymm}
  \end{equation}
  in distribution, where
  \begin{align*}
    \Lambda_{\rm all} \enspace &= \enspace \Gamma_{\rm all}^{-1} \; \Sigma_{\rm all} \; \Gamma_{\rm all}^{\rm -T}; \\
    \Sigma_{\rm all}  \enspace &= \enspace \cov\left\{\begin{matrix} \zeta_{X*}(Z,\Omega) \\ \zeta_{G*}(Z,\Omega) \end{matrix}\right\} \; = \enspace \cov\left\{\begin{matrix} \zeta_{X}(Z,\Omega) \\ \zeta_{G}(Z,\Omega) \end{matrix}\right\}.
  \end{align*}
}\end{Thm}

The proof of \cref{thm:AsympNormSymm} follows directly from the proofs of \cref{eq:asymp_X,eq:asymp_G} and the properties of M-estimators $\wh{\Omega}_X$ and $\wh{\Omega}_G$.

\begin{Rem}\label{rem:BootstrapCov}
  In \cref{sec:Symm}, we constructed a linear model from \cref{eq:AsympNormSymm} and used generalized least squares to calculate $\wh{\Omega}_{\rm Symm}$.  The asymptotic properties of GLS estimators inform us that as $n \to \infty$, 
  \begin{equation*}
    n^{1/2}(\wh{\Omega}_{\rm Symm}-\Omega) \to \Normal\{0,({\cal X}\trans \Lambda^{-1}_{\rm all}{\cal X})^{-1}\}.
  \end{equation*}
  
  In practice, $\wh\Omega_{X}$ and $\wh\Omega_{G}$ are highly correlated, which slows convergence to the asymptotic covariance matrix.  Asymptotic estimates of standard errors proved unreliable in simulations, and are not recommended.  Instead, we estimate $\cov(\wh{\Omega}_{\rm Symm})$ using a balanced bootstrap, where cases and controls are resampled separately, thus maintaining their respective sample sizes.
\end{Rem}

\subsection{Rare Diseases When $\pi_1$ is Unknown} \label{sec:RareDisease}

Due to its sampling efficiency, the case-control design is typically used to study relatively rare diseases.  If the true disease rate in the source population is unknown, it is common to assume that $\pi_1 \approx 0$ \citep{Piegorsch1994,modan2001parity,lin2006likelihood,kwee2007simple}.  Under this rare disease assumption, $\wh\Omega_{X}$ and $\wh\Omega_{G}$ converge not to $\Omega$, but to $\Omega_{X*}$ and $\Omega_{G*}$, the solutions to their respective score equations with $\pi_1=0$.  Using estimates of $\Omega_{X*}$ and $\Omega_{G*}$ to calculate $\wh{\Omega}_{\rm Symm}$ runs the risk of introducing bias, but in practice the small potential bias is typically inconsequential unless the sample size is very large and standard errors unusually small.  Examples in \cref{sec:Misspec} of the \Supp{} demonstrate that under the rare disease approximation, coverage intervals remain near nominal until the true disease rate exceeds 8\%.

{\subsection{Asymptotic efficiency}
In this section, we discuss the relative efficiency of the proposed estimator $\widehat{\Omega}_{\rm Symm}$ with SPMLE estimator $\widehat{\Omega}_{X}$. Recall that $\Omega = (\kappa, \bbeta\trans)\trans$,  and $\bbeta = (\beta_1,...,\beta_p)$ is the parameter of interest. Since both estimators are consistent, our goal is to show ${\rm var}(\beta_{\rm Symm, j})\leq{\rm var}(\beta_{\rm X, j})$ for $j =1,...,p$. We start with  their asymptotic covariance matrix $({\cal{X}}\trans \Lambda_{\rm all}^{-1}{\cal{X}})^{-1}$ and $\Gamma^{-1}_X{\rm cov}\{\xi_{X}(Z,\Omega)\}\Gamma_X^{\rm -T}$. Recall that
\bse 
\Lambda_{\rm all}^{-1}
&=& \begin{pmatrix}\Gamma^{-1}_X{\rm cov}\{\xi_X(Z,\Omega)\}\Gamma^{\rm -T}_X&\Gamma^{-1}_X{\rm cov}\{\xi_X(Z,\Omega), \xi_G(Z,\Omega)\}\Gamma^{\rm -T}_G\\
\Gamma^{-1}_G{\rm cov}\{\xi_G(Z,\Omega), \xi_X(Z,\Omega)\}\Gamma^{\rm -T}_X& \Gamma^{-1}_G{\rm cov}\{\xi_G(Z,\Omega)\}\Gamma^{\rm -T}_G\\
\end{pmatrix}^{-1}.\\
\ese 
To simplify the notations, we denote $A = \Gamma^{-1}_X{\rm cov}\{\xi_X(Z,\Omega)\}\Gamma^{\rm -T}_X$, $B= \Gamma^{-1}_X{\rm cov}\{\xi_X(Z,\Omega), \xi_G(Z,\Omega)\}\Gamma^{\rm -T}_G$, and $D = \Gamma^{-1}_G{\rm cov}\{\xi_G(Z,\Omega)\}\Gamma^{\rm -T}_G$. According to \cite[Theorem 2.1]{lu2002inverses},  then we can further write $\Lambda^{-1}_{all}$ as
\bse 
\Lambda_{\rm all}^{-1}&=&\begin{pmatrix}A&B\\
B\trans&D\end{pmatrix}^{-1}= \begin{pmatrix}
A^{-1}+A^{-1}B(D-B\trans A^{-1}B)^{-1}B\trans A^{-1}
& -A^{-1}B(D-B\trans A^{-1}B)^{-1}\\
-(D-B\trans A^{-1}B)^{-1}B\trans A^{-1}& (D-B\trans A^{-1}B)^{-1}
\end{pmatrix}.
\ese
Thus, 
\bse
{\cal{X}}\trans \Lambda_{\rm all}^{-1}{\cal{X}}&=& \begin{pmatrix} I_p&I_p\end{pmatrix}
\begin{pmatrix}
A^{-1}+A^{-1}B(D-B\trans A^{-1}B)^{-1}B\trans A^{-1}
& -A^{-1}B(D-B\trans A^{-1}B)^{-1}\\
-(D-B\trans A^{-1}B)^{-1}B\trans A^{-1}& (D-B\trans A^{-1}B)^{-1}
\end{pmatrix} \begin{pmatrix} I_p\\I_p\end{pmatrix}\\
&=& A^{-1}+A^{-1}B(D-B\trans A^{-1}B)^{-1}B\trans A^{-1}
 -A^{-1}B(D-B\trans A^{-1}B)^{-1}\\
 && \hskip 3mm -(D-BB\trans A^{-1}B)^{-1}B\trans A^{-1}+ (D-B\trans A^{-1}B)^{-1}\\
 &=& A^{-1} + \left\{(D-B\trans A^{-1}B)^{-1/2}(B\trans A^{-1} -I_p)\right\}\trans \left\{(D-B\trans A^{-1}B)^{-1/2}(B\trans A^{-1} -I_p)\right\}.
\ese 
To further simplify the notations, we denote $\left\{(D-B\trans A^{-1}B)^{-1/2}(B\trans A^{-1} -I_p)\right\}$ as $C$, and write ${\cal{X}}\trans \Lambda_{\rm all}^{-1}{\cal{X}}=A^{-1}+C\trans C$. Thus,
\bse 
{\rm cov}(\widehat{\Omega}_{\rm G}) - {\rm cov}(\widehat{\Omega}_{\rm Symm}) &=& (A^{-1}+C\trans C)^{-1}-A\\
&=& A - \{A - AC\trans (I+CAC\trans)^{-1}CA \}\\
&=& AC\trans (I+CAC\trans)^{-1}CA\\
&\geq&0.
\ese 
We observe that the matrix $AC\trans (I+CAC\trans)^{-1}CA$ is positive semi-definite, and its diagonal elements are non-negative. Hence, we conclude that ${\rm var}(\beta_{\rm Symm, j}) \leq {\rm var}(\beta_{\rm X, j}), j = 1,...,p$. In practice, ${\rm var}(\beta_{\rm Symm, j}) = {\rm var}(\beta_{\rm X, j})$ when  $D = B\trans A^{-1} B$. According to the mathematical symmetry , we get similar results that ${\rm var}(\beta_{\rm Symm, j}) \leq {\rm var}(\beta_{\rm G, j}), j = 1,...,p$, and ${\rm var}(\beta_{\rm Symm, j}) = {\rm var}(\beta_{\rm G, j})$ when $A = B\trans D^{-1} B$.

As we discussed at the beginning of this paper, though the estimator $\widehat{\Omega}_X$ is proposed in the method of Stalder et al., $\widehat{\Omega}_G$ is an alternative choice according to the fact of mathematical symmetry. Choosing $\widehat{\Omega}_X$ or $\widehat{\Omega}_G$ can be tricky and require prior information in practice. However, through the above illustration, we show that the proposed estimator $\widehat{\Omega}_{\rm Symm}$ is guaranteed to be more efficient than, or at least as efficient as, either $\widehat{\Omega}_X$ or $\widehat{\Omega}_G$, without worrying about the arbitrary choice between $\widehat{\Omega}_X$ and $\widehat{\Omega}_G$.  
}

\section{Simulations} \label{sec:Simulations}

\subsection{Scenario} \label{sec:SimScenario}

To investigate the performance of the Symmetric Combination Estimator, we adopt the same simulation settings as reported in \citet{Stalder2017}.  Environmental variable $X$ is binary with population frequency 0.5, and $G$ consists of five correlated single nucleotide polymorphisms (SNPs).  The SNPs follow a trinomial distribution in Hardy-Weinberg equilibrium, wherein SNP $G_j$ takes values $(0, 1, 2)$ with probabilities $\{(1-p_j)^2, \enspace 2p_j(1-p_j), \enspace p_j^2\}$, respectively.

To generate correlated SNPs, we first simulated a 5-variate normal random variable with mean 0 and covariance between the $j$th and $k$th components equal to $0.7^{|j-k|}$.  We then trichotomized the variates with appropriate thresholds so that the frequency of 0, 1, and 2 followed Hardy-Weinberg equilibrium with minor allele frequencies $(p_1, p_2, p_3, p_4, p_5) = (0.1, 0.3, 0.3, 0.3, 0.1)$.

Disease status was simulated according to the risk model $H\{\alpha_0 + m(G,X,\bbeta)\}$, with $m(G,X,\beta) = G\trans \beta_G+X\beta_X + (GX)\trans\beta_{GX}$.  Here $\beta_G = \{\log(1.2), \log(1.2), 0, \log(1.2), 0\}$, $\beta_X = \log(1.5)$, and $\beta_{GX}=\{\log(1.3), 0, 0, \log(1.3), 0\}$.  We set the logistic intercept $\alpha_0 = -4.165$ to yield a population disease rate $\pi_1 = 0.03$.

A sample of 1000 cases and 1000 controls was drawn from the simulated population, and parameters were estimated using logistic regression, the SPMLE of \citet{Stalder2017}, the Symmetric Combination Estimator with known $\pi_1$, and the Symmetric Combination Estimator with a rare disease approximation.  Standard error estimates for both logistic regression and the SPMLE were based on asymptotic theory, while those for the Symmetric Combination Estimator were calculated using 200 balanced bootstrap samples, as described in \cref{rem:BootstrapCov}.

\subsection{Results} \label{sec:SimResults}

\Cref{tab:sim1} presents the results of 1000 simulations comparing standard logistic regression, the SPMLE proposed by \citeauthor{Stalder2017} with known $\pi_1$, the SPMLE proposed by \citeauthor{Stalder2017} using the rare disease approximation, our proposed Symmetric Combination Estimator with known $\pi_1$, and the Symmetric Combination Estimator using the rare disease approximation.  Standard error estimates for logistic regression and the SPMLE were calculated using asymptotic theory, and the standard error estimates for both versions of the Symmetric Combination Estimator were calculated using 200 bootstrap samples as described in \cref{rem:BootstrapCov}.

The Symmetric Combination Estimator, both with known $\pi_1$ and when using the rare disease approximation, shows negligible bias and has coverage percentages near the nominal level.  Like the SPMLE, both versions of our Symmetric Combination Estimator provide slightly more than 25\% improvement in mean squared error efficiency over ordinary logistic regression for the main effect of $X$.

More impressively, our estimator nearly doubles the mean squared error efficiency of logistic regression for the main effects of $G$, and nearly triples the mean squared error efficiency for the interaction terms.  This is a marked improvement even over the performance of the SPMLE, and it is accomplished without modeling the distribution of either $G$ or $X$.

\begin{table}[!ht]
\ttabbox
  {\caption{\baselineskip=12pt Results of 1000 simulations as described in \cref{sec:SimScenario}, comparing the bias, coverage, and efficiency of four estimators: ordinary logistic regression, the SPMLE of \citeauthor{Stalder2017} with known $\pi_1$, our proposed Symmetric Combination Estimator with known $\pi_1$, and the Symmetric Combination Estimator using the rare disease approximation} \label{tab:sim1}}
  {\begin{tabular}{ | l | ll l l  l | l | l  l  l  l  l| }
    \hline
    &$\beta_{G1}$ &$\beta_{G2}$&$\beta_{G3}$ & $\beta_{G4}$ & $\beta_{G5}$ & $\beta_{X}$ & $\beta_{XG1}$ & $\beta_{XG2}$ & $\beta_{XG3}$ &$\beta_{XG4}$ & $\beta_{XG5}$  \\ \hline
    True & 0.18&	0.18&	0.00&	0.18&	0.00	&0.41&	0.26&	0.00&	0.00&	0.26&	0.00 \\
    \hline
    \multicolumn{12}{|c|}{}\\[-.80em]\multicolumn{12}{|c|}{Logistic Regression}\\
    \hline
    Bias & 0.01&	0.00&	0.00&0.00&	-0.01	&0.00&	0.00&	0.00&	0.00	&0.00&	0.00 \\
    CI(\%) & 95.2&	95.5&	94.4&	94.7&	95.3&	95.8&	94.5	&95.9&	94.7&	94.6&	95.3 \\
    \hline
    \multicolumn{12}{|c|}{}\\[-.80em]
    \multicolumn{12}{|c|}{SPMLE, known $\pi_1$} \\
      \hline
      Bias      & 0.01 & 0.00  & 0.00 & 0.00 & -0.01 & 0.00 & 0.00  & 0.00 & 0.00 & 0.00  & 0.01 \\
      CI(\%)    & 95.4 & 95.8  & 94.8 & 96.1 & 96.3  & 95.6 & 94.6  & 96.0 & 94.3 & 95.6  & 95.1 \\
      MSE Eff   & 1.32 & 1.25  & 1.26 & 1.32 & 1.30  & 1.27 & 2.08  & 1.78 & 1.88 & 1.95  & 2.12 \\
      \hline
      \multicolumn{12}{|c|}{}\\[-.80em]
      \multicolumn{12}{|c|}{SPMLE, rare} \\
      \hline
      Bias      & 0.02 & 0.00  & 0.00 & 0.01 & -0.01 & 0.02 & -0.02 & 0.00 & 0.00 & -0.01 & 0.01 \\
      CI(\%)    & 95.0 & 95.7  & 94.8 & 95.9 & 96.4  & 95.5 & 94.4  & 96.0 & 94.7 & 95.4  & 95.7 \\
      MSE Eff   & 1.31 & 1.26  & 1.27 & 1.32 & 1.30  & 1.26 & 2.18  & 1.89 & 2.01 & 2.06  & 2.27 \\
      
    \hline
    \multicolumn{12}{|c|}{}\\[-.80em]\multicolumn{12}{|c|}{Symmetric Combination Estimator, known $\pi_1$}\\
    \hline 
    Bias &0.00	&-0.03&	0.00&	0.00&	-0.01&	0.01&	-0.03&	0.02&	0.00&	-0.02&	0.01 \\
    CI$^*$(\%) & 96.7&	95.7&	96.7&	96.5&	97.8&	95.4&	94.8&	96.7&	96.2&	96.6&	97.2 \\
    MSE Eff   & 1.92 & 1.71  & 2.00 & 1.83 & 2.05  & 1.31 & 2.84  & 2.51 & 2.99 & 2.68  & 3.34 \\
      \hline
       \multicolumn{12}{|c|}{}\\[-.80em]
      \multicolumn{12}{|c|}{Symmetric Combination Estimator, rare} \\
      \hline
      Bias      & 0.01 & -0.02 & 0.00 & 0.01 & -0.01 & 0.02 & -0.05 & 0.02 & 0.00 & -0.03 & 0.00 \\
      CI$^*$(\%)& 96.4 & 95.7  & 95.7 & 96.3 & 98.1  & 94.9 & 94.0  & 97.0 & 96.4 & 95.5  & 97.5 \\
      MSE Eff   & 1.86 & 1.71  & 1.92 & 1.78 & 1.95  & 1.27 & 2.75  & 2.66 & 3.08 & 2.69  & 3.58 \\
      
    \hline
  \end{tabular}
  \floatfoot{\baselineskip=11pt \emph{CI}: coverage of a 95\% nominal confidence interval, calculated using asymptotic standard error.  \emph{CI$^*$}: coverage of a 95\% nominal confidence interval, calculated using 200 bootstrap samples.  \emph{MSE Eff}: mean squared error efficiency when compared to logistic regression.}}
\end{table}

\subsection{Further Simulations} \label{sec:SimFurther}

Further simulations were conducted with multiple correlated SNPs and a binary environmental risk factor, but with changes to the number of SNPs (3 or 8), the population disease rate (1\% or 5\%), or the sample size (500 or 3000 cases \& controls).  All such simulations yielded results similar to those in \cref{tab:sim1} with regards to coverage, efficiency gains, and unbiasedness, and are thus not reported.

\Cref{sec:AddlSims} of the \Supp{} contains the results of simulations examining the behavior of the Symmetric Combination Estimator in a variety of settings.  \Cref{sec:sim1ALL} contains an unabridged version of \cref{tab:sim1} that includes the SPMLE\_G ($\wh{\Omega}_G$) and the Composite Likelihood Estimator, neither of which approach the MSE efficiency of the Symmetric Combination Estimator.  \Cref{sec:Misspec} presents the results of simulations with misspecified population disease rate; we found the Symmetric Combination Estimator fairly robust to the misspecification of the disease rate.  \Cref{sec:ViolAssump} contains simulation studies examining the robustness of our method with respect to violations of the gene-environment independence assumption.  Those simulations demonstrate that there will be bias in the estimated interaction parameter between a specific gene and a correlated environmental variable, but the rest of the parameter estimates continue unbiased, and the average mean squared error for all $G{*}X$ interactions can still be substantially lower than that obtained from prospective logistic regression.  \Cref{sec:AltDist} presents the results of simulations with different distributions for $G$ and $X$.

\section{Data Analysis}\label{sec:DataAnalysis}

\subsection{Data}\label{sec:DataData}

Here we apply the proposed methodology to a case-control study of breast cancer.  This case-control sample is taken from a large prospective cohort at the National Cancer Institute: the Prostate, Lung, Colorectal and Ovarian cancer screening trial \citep{canzian2010comprehensive}.  This cohort enrolled 64,440 non-Hispanic, white women aged 55 to 74, of whom 3.72\% developed breast cancer \citep{pfeiffer2013risk}.  The case-control study analyzed here consists of 658 cases and 753 controls.

Each of the 1411 subjects was genotyped for 21 SNPs that have been previously associated with breast cancer based on large genome-wide association studies.  These SNPs were weighted by their log-odds-ratio coefficients and summed to define a polygenic risk score.  A scaled version of this polygenic risk score, with mean zero and standard deviation one, was used as the genetic risk factor $G$.  The individual SNPs and their coefficients can be found in \cref{sec:PRSweights} of the \Supp{}.

Early menarche is a known risk factor for breast cancer \citep{anderson2007estimating}, and environmental variable $X$ is a binary indicator of whether the age at menarche is less than 14.  The interaction between age at menarche and genetic breast cancer risk is a topic of interest, but the power to detect such interactions in previous studies has been limited \citep{gail2008discriminatory}.

The model fitted is $\pr(D = 1) = H(\beta_0+\beta_G G + \beta_X X + \beta_{GX} GX)$.  While $\pi_1$ is known in this population, we apply our method using both the known disease rate and the rare disease approximation.

\subsection{Verifying Gene-Environment Independence}\label{sec:DataIndepend}

Before applying our approach, we performed analyses to check the assumption of gene-environment independence in the population.  Using the 753 controls, we ran a $t$-test of the polygenic risk score against the levels of $X$.  The $p$-value was 0.91, indicating no evidence of correlation between $G$ and $X$.  We also ran chi-squared tests for each of the 21 individual genes and found no significant association after controlling the false discovery rate: the minimum $q$-value was 0.09.

We also checked for correlation, known as linkage disequilibrium, between the 21 SNPs used to create the polygenic risk score and 32 SNPs known to influence age at menarche \citep{elks2010menarche}.  Using phased haplotypes from subjects of European descent from \emph{1000 Genomes} \citep{1000Genomes2015} and \emph{HapMap} \citep{gibbs2003international}, we were able to analyze 651 of the 672 possible linkages, and no evidence of linkage disequilibrium was found: the maximum $R^2$ was 0.1 and the minimum q-value was 0.85.  Finally, a recent study of breast cancer susceptibility loci examined the relationship between age at menarche and 10 of the 21 SNPs used to create our polygenic risk score, none of which were found to influence age at menarche \citep{andersen2014susceptibility}.

\subsection{Results}\label{sec:DataResults}

\renewcommand\FBaskip{14pt}
\floatbox[\captop]{table}[0.9\hsize][][]
{
  \caption{\baselineskip=12pt Results of the analysis of the Prostate, Lung, Colorectal and Ovarian cancer screening trial data} \label{tab:PLCO}
}{
  \begin{tabular}{|l|lll|}
    \hline
    & $\beta_G$ &$\beta_X$   &$\beta_{GX}$ \\
    \hline
    {Logistic Regression}&&& \\
    \qquad\qquad Estimate                   & 0.539 & 0.124 &-0.242 \\*[-.40em]
    \qquad\qquad Standard Error (asymptotic)& 0.117 & 0.128 & 0.133 \\*[-.40em]
    \qquad\qquad p-value (asymptotic)       &$<1e$-4& 0.331 & 0.068 \\
    \hline
    {Symmetric Combination, known $\pi_1=3.72\%$}&&& \\
    \qquad\qquad Estimate                   & 0.495 & 0.093 &-0.215 \\*[-.40em]
    \qquad\qquad Standard Error (bootstrap) & 0.094 & 0.133 & 0.089 \\*[-.40em]
    \qquad\qquad p-value (bootstrap)        &$<1e$-4& 0.484 & 0.016 \\
    \hline
    {SPMLE, known $\pi_1=3.72\%$}&&& \\
    \qquad\qquad Estimate                   & 0.587&0.127&-0.274 \\*[-.40em]
    \qquad\qquad Standard Error (asymptotic) & 0.103& 0.128 & 0.108 \\*[-.40em]
    \qquad\qquad p-value (asymptotic)        &$<1e$-4& 0.321 & 0.011 \\
    \hline
    {Symmetric Combination, rare disease approximation}&&& \\
    \qquad\qquad Estimate                   & 0.538 & 0.116 &-0.237 \\*[-.40em]
    \qquad\qquad Standard Error (bootstrap) & 0.089 & 0.124 & 0.099 \\*[-.40em]
    \qquad\qquad p-value (bootstrap)        &$<1e$-4& 0.352 & 0.016 \\
    \hline
      {SPMLE, rare disease approximation}&&& \\
    \qquad\qquad Estimate                   & 0.590& 0.129 &-0.269 \\*[-.40em]
    \qquad\qquad Standard Error (asymptotic) & 0.104 & 0.128 & 0.106 \\*[-.40em]
    \qquad\qquad p-value (asymptotic)        &$<1e$-4&0.315 & 0.012 \\
    \hline
  \end{tabular}
    \floatfoot{\baselineskip=11pt $\beta_{G}$ and $\beta_{X}$ are the main effects for the polygenic risk score $G$ and the environmental variable $X$ (age at menarche $<$ 14), and $\beta_{GX}$ is the gene-environment interaction.}
}

\Cref{tab:PLCO} presents the results of our analysis with known $\pi_1$ and under a rare disease approximation.  In both cases, standard errors for the Symmetric Combination Estimator were calculated using 500 bootstrap samples.  The two estimates yield very similar results, indicating that a valid analysis can be conducted even if $\pi_1$ is not known. The p-value for SPMLE estimates are smaller, due to the larger estimates compared to Symmetric Combination Estimator. However, the Symmetric Combination Estimator often has smaller standard error, especially for the gene-environment interaction term.

The polygenic risk score was strongly associated with breast cancer status of the women in the study, which is to be expected given that each of its component SNPs has a known association with breast cancer risk.  Standard logistic regression analysis provides some indication of an interaction between the polygenic risk score and age at menarche, but the result is not statistically significant at the 0.05 level.  Using the assumption of gene-environment independence in the population, the Symmetric Combination Estimator finds stronger evidence of this interaction.  The improved power to detect this interaction is due to the much smaller standard error estimates of the Symmetric Combination Estimators.  Using logistic regression, the estimated standard error of $\beta_{GX}$ is 49\% larger than with our method, indicating a variance increase of 121\% (when applying the rare disease approximation, the variance increase is 81\%).

\section{Discussion and Extensions}\label{sec:Discussion}

Researchers investigating gene-environment interactions in case-control studies have traditionally had two broad options for analysis: standard logistic regression, which is flexible but has low power to detect interactions, or retrospective methods, which lack flexibility but offer improved efficiency by exploiting the assumption of gene-environment independence.  Improved understanding of genetic risk factors has led to the need for efficient estimators that can model complex gene-environment interactions.  \Citet{Stalder2017} proposed a retrospective profile method that exploits the assumption of gene-environment independence while treating the genetic and environmental variables nonparametrically.  By obviating the need for a parametric model of genotype distributions, their method is well suited for the analysis of multimarker genetic data and polygenic risk scores. 

We observed the mathematical symmetry of the genetic and environmental variables and discovered the sub-optimal efficiency gain in \citet{Stalder2017}. Hence, we proposed an improvement to the method of \citet{Stalder2017} that substantially increases the efficiency of the estimates with modest computational cost and no additional assumptions, making it applicable anywhere that the method of \citeauthor{Stalder2017} can be used.  Simulations under a variety of scenarios demonstrate a consistent improvement in mean squared error efficiency over the method of \citet{Stalder2017} and logistic regression on the estimation of both main effects and gene-environment interaction terms.  Our methods are implemented in the R package \texttt{caseControlGE}, freely available at \url{github.com/alexasher/caseControlGE}.

The proposed Symmetric Combination Estimator places no distributional assumptions on the genetic or environmental variables, but it does rely on three assumptions.  The first assumption, that the logistic risk model $H\{\alpha_0 + m(G,X,\bbeta)\}$ is known up to parameters $\alpha_0$ and $\bbeta$, is minimally restrictive because a flexible function, such as a function of B-splines, can be defined for $m(G,X,\bbeta)$.  The second assumption, that $\pi_1$ is known or can be well estimated, can be relaxed by using the rare disease approximation of \cref{sec:RareDisease}.  Even if the true disease rate is not rare, the Symmetric Combination Estimator is generally robust to the misspecification of $\pi_1$, as demonstrated in \cref{sec:Misspec} of the \Supp{}.

The final assumption is gene-environment independence in the source population.  In \cref{sec:ViolAssump} of the \Supp{}, we present the results of simulations demonstrating that bias is introduced in the estimated interaction parameter between correlated genetic and environmental variables, but the rest of the parameter estimates are unbiased.  We recommend that researchers verify gene-environment independence before applying the Symmetric Combination Estimator, as we did in \cref{sec:DataIndepend}.  To relax the gene-environment independence assumption, it should be straightforward to adapt the Symmetric Combination Estimator to the case where $G$ and $X$ are conditionally independent within the strata of an observed factor, as demonstrated in the \Supp{} of \citet{Stalder2017}.  If suitable strata cannot be found, another possibility is to construct an empirical Bayes-type shrinkage estimator like that of \citet{mukherjee2008exploiting}, which would shrink the estimate from standard logistic regression back to the Symmetric Combination Estimator when the gene-environment independence assumption is valid.

\section*{Supplementary Material}
The \Supp\ includes methodology and theory for the Composite Likelihood Estimator, the unabridged version of \cref{tab:sim1} with all estimators, and additional simulation results for model robustness when the disease rate is misspecified or the gene-environment independence assumption is violated.

\section*{Acknowledgement}

We would like to thank Dr. Raymond J. Carroll and Dr. Yanyuan Ma for sharing their valuable insights about case-control studies. The research of Wang and Asher was supported by a grant from the National Cancer Institute (U01-CA057030).

\bibliographystyle{biomAbhra}
\bibliography{TWAA_10-10-20}

\newcommand{\Appendix}{\appendix\def\thesection{Appendix~\Alph{section}}\def\thesubsection{\Alph{section}.\arabic{subsection}}}
\clearpage\pagebreak\newpage
\pagestyle{fancy}
\fancyhf{}
\rhead{\bfseries\thepage}
\lhead{\bfseries NOT FOR PUBLICATION SUPPLEMENTARY MATERIAL}
\begin{center}
	{\LARGE{\bf Supplementary Material }}
\end{center}
\vskip 2mm
\setcounter{figure}{0}
\setcounter{equation}{0}
\setcounter{Thm}{0}
\setcounter{page}{1}
\setcounter{table}{1}
\setcounter{section}{0}
\renewcommand{\thefigure}{S.\arabic{figure}}
\renewcommand{\theequation}{S.\arabic{equation}}
\renewcommand{\theThm}{S.\arabic{Thm}}
\renewcommand{\thesection}{S.\arabic{section}}
\renewcommand{\thesubsection}{S.\arabic{section}.\arabic{subsection}}
\renewcommand{\thepage}{S.\arabic{page}}
\renewcommand{\thetable}{S.\arabic{table}}

\section{Composite Likelihood Estimator}\label{sec:CompositeLik}

The estimated composite profile likelihood is just the average of the two symmetric profile likelihoods
\bse 
   \wh{\cal L}_{\rm CL}(\Omega) &=& \{\wh{\cal L}_{X}(\Omega) + \wh{\cal L}_{G}(\Omega)\}/2 \\
                              &=& \sumi \log\{S(D_i,G_i,X_i,\Omega)\} - 0.5\sumi \log\{\wh{R}_{X}(G_i,\Omega)\}- 0.5\sumi \log\{\wh{R}_{G}(X_i,\Omega)\}.
\ese 
The estimated score function is thus the average of the two symmetric score functions
\bse
\wh{{\cal S}}_{\rm CL}(\rm \Omega) &=& (\wh{{\cal S}}_{X}(\rm \Omega) + \wh{{\cal S}}_{G}(\rm \Omega))/2 \\
                                 &=& n^{-1/2}\sum_{i=1}^{n}\left\{ \frac{S_{\Omega}(D_i,G_i,X_i,\Omega)}{S(D_i,G_i,X_i,\Omega)} -
\frac{1}{2}\frac{\wh{R}_{X \Omega}(X_i,\Omega)}{\wh{R}_{X}(X_i,\Omega)}-\frac{1}{2}\frac{\wh{R}_{G \Omega}(G_i,\Omega)}{\wh{R}_{G}(G_i,\Omega)}\right\}. \label{eq:composite_score}
\ese
Estimate $\wh{\Omega}_{\rm CL}$ is calculated by solving $\wh{{\cal S}}_{\rm CL}(\Omega)=0$, or equivalently, maximizing $\wh{\cal L}_{\rm CL}(\Omega)$.

Following the notation defined previously, we sum \cref{eq:asymp_X,eq:asymp_G} together instead of stacking them as in \cref{thm:AsympNormSymm}. 
\begin{Thm}\label{thm:AsympNormComposite}{
  Suppose that $0 <  \displaystyle\lim_{n\to\infty} n_d/n<1$, and $\pi_1$ is known. Then
  \begin{equation*}
    n^{1/2}(\wh{\Omega}_{\rm CL}- \Omega) = -(\Gamma_{X} + \Gamma_{G})^{-1} n^{-1/2} \sumi \{\zeta_{X*}(Z_i,\Omega)+\zeta_{G*}(Z_i,\Omega)\}  + o_p(1).
 \end{equation*}
  To calculate the asymptotic variance, write
  \begin{align*}
    \Sigma_{\rm all}  \enspace &= \enspace \left[\begin{matrix} \Sigma_{XX} & \Sigma_{XG}\\ \Sigma_{GX} & \Sigma_{GG} \end{matrix}\right] \; = \enspace \cov\left\{\begin{matrix} \zeta_{X*}(Z_i,\Omega) \\ \zeta_{G*}(Z_i,\Omega) \end{matrix}\right\} \; = \enspace \cov\left\{\begin{matrix} \zeta_{X}(Z_i,\Omega) \\ \zeta_{G}(Z_i,\Omega) \end{matrix}\right\};\\
    \Sigma_{XX} \enspace &= \enspace \sumd (n_d/n) \cov\{\zeta_{X*}(Z,\Omega) \vert D=d\} \enspace = \enspace \sumd (n_d/n) \cov\{\zeta_{X}(Z,\Omega) \vert D=d\};  \\
    \Sigma_{GG} \enspace  &= \enspace \sumd (n_d/n) \cov\{\zeta_{G*}(Z,\Omega) \vert D=d\} \enspace = \enspace \sumd (n_d/n) \cov\{\zeta_{G}(Z,\Omega) \vert D=d\};  \\
    \Sigma_{XG}  \enspace &= \enspace \sumd (n_d/n) \cov\{\zeta_{X*}(Z,\Omega),\zeta_{G*}(Z,\Omega) \vert D=d\} \\
                &= \enspace \sumd (n_d/n) \cov\{\zeta_{X}(Z,\Omega),\zeta_{G}(Z,\Omega) \vert D=d\} \enspace = \enspace \Sigma_{XG}\trans .
  \end{align*}
  Since the $Z_i$ are independent and $E\{\zeta_{X*}(Z_i,\Omega)\vert D_i\}=E\{\zeta_{G*}(Z_i,\Omega)\vert D_i\} = 0$, then
  \bse 
    n^{1/2}(\wh{\Omega}_{\rm CL}- \Omega) &\to& \Normal(0,\Lambda_{\rm CL}); \\
    \Sigma_{\rm CL} &=& \sumd (n_d/n) \cov\{\zeta_{X*}(Z_i,\Omega)+\zeta_{G*}(Z_i,\Omega)\} \\
    &=& \Sigma_{XX}+\Sigma_{GG}+\Sigma_{XG}+\Sigma_{GX}; \\
    \Lambda_{\rm CL} &=& (\Gamma_{X} + \Gamma_{G})^{-1} \Sigma_{\rm CL} \{(\Gamma_{X} + \Gamma_{G})^{-1}\}\trans.
  \ese
}\end{Thm}
The proof of \cref{thm:AsympNormComposite} follows directly from the proofs of \cref{eq:asymp_X,eq:asymp_G} and the properties of M-estimators $\wh{\Omega}_X$ and $\wh{\Omega}_G$.

\section{Additional Simulations}\label{sec:AddlSims}

\subsection{Unabridged version of \cref{tab:sim1} from \cref{sec:Simulations}} \label{sec:sim1ALL}

\Cref{tab:sim1} in \cref{sec:Simulations} of the main paper reports the results of five estimators: logistic regression, the SPMLE with known $\pi_1$, the SPMLE using the rare disease approximation, our Symmetric Combination Estimator with known $\pi_1$, and our Symmetric Combination Estimator using the rare disease approximation.  \Cref{tab:sim1ALL} presents the results of \emph{all} estimators in 1000 simulations under the simulation settings of \cref{sec:SimScenario}.  In addition to logistic regression, four retrospective methods are presented: the SPMLE ($\wh{\Omega}_X$), the SPMLE\_G ($\wh{\Omega}_G$), the Composite Likelihood Estimator ($\wh{\Omega}_{\rm CL}$), and the Symmetric Combination Estimator ($\wh{\Omega}_{\rm Symm}$).  Each retrospective estimator was calculated under two conditions: with known $\pi_1$, and with unknown $\pi_1$ using the rare disease approximation.

We see that the rare disease approximation of each retrospective estimator closely matches the version calculated with known $\pi_1$.  The efficiency of the Composite Likelihood Estimator is equivalent to that of the SPMLE and its symmetric counterpart, the SPMLE\_G.  The Symmetric Combination Estimator stands out as markedly more efficient than the other estimators.

\begin{table}[H]
\ttabbox
  {\caption{\baselineskip=12pt Results of 1000 simulations as described in \cref{sec:SimScenario}, comparing the bias, coverage, and efficiency of all estimators} \label{tab:sim1ALL}}
  {\begin{tabular}{ | l | l l l l l | l | l l l l l | }
      \hline
      &$\beta_{G1}$ &$\beta_{G2}$&$\beta_{G3}$ & $\beta_{G4}$ & $\beta_{G5}$ & $\beta_{X}$ & $\beta_{XG1}$ & $\beta_{XG2}$ & $\beta_{XG3}$ &$\beta_{XG4}$ & $\beta_{XG5}$ \\
      \hline
      True      & 0.18 & 0.18  & 0.00 & 0.18 & 0.00  & 0.41 & 0.26  & 0.00 & 0.00 & 0.26  & 0.00 \\
      \hline
      \multicolumn{12}{|c|}{Logistic Regression} \\
      \hline
      Bias      & 0.01 & 0.00  & 0.00 & 0.00 & -0.01 & 0.00 & 0.00  & 0.00 & 0.00 & 0.00  & 0.00 \\
      CI(\%)    & 95.2 & 95.5  & 94.4 & 94.7 & 95.3  & 95.8 & 94.5  & 95.9 & 94.7 & 94.6  & 95.3 \\
      \hline
      \multicolumn{12}{|c|}{SPMLE, known $\pi_1$} \\
      \hline
      Bias      & 0.01 & 0.00  & 0.00 & 0.00 & -0.01 & 0.00 & 0.00  & 0.00 & 0.00 & 0.00  & 0.01 \\
      CI(\%)    & 95.4 & 95.8  & 94.8 & 96.1 & 96.3  & 95.6 & 94.6  & 96.0 & 94.3 & 95.6  & 95.1 \\
      MSE Eff   & 1.32 & 1.25  & 1.26 & 1.32 & 1.30  & 1.27 & 2.08  & 1.78 & 1.88 & 1.95  & 2.12 \\
      \hline
      \multicolumn{12}{|c|}{SPMLE, rare} \\
      \hline
      Bias      & 0.02 & 0.00  & 0.00 & 0.01 & -0.01 & 0.02 & -0.02 & 0.00 & 0.00 & -0.01 & 0.01 \\
      CI(\%)    & 95.0 & 95.7  & 94.8 & 95.9 & 96.4  & 95.5 & 94.4  & 96.0 & 94.7 & 95.4  & 95.7 \\
      MSE Eff   & 1.31 & 1.26  & 1.27 & 1.32 & 1.30  & 1.26 & 2.18  & 1.89 & 2.01 & 2.06  & 2.27 \\
      \hline
      \multicolumn{12}{|c|}{SPMLE\_G, known $\pi_1$} \\
      \hline
      Bias      & 0.01 & 0.00  & 0.00 & 0.00 & -0.01 & 0.00 & 0.00  & 0.00 & 0.00 & 0.00  & 0.01 \\
      CI(\%)    & 95.0 & 95.8  & 94.8 & 96.1 & 96.3  & 95.2 & 94.8  & 95.5 & 94.1 & 95.6  & 95.6 \\
      MSE Eff   & 1.35 & 1.27  & 1.29 & 1.34 & 1.33  & 1.28 & 2.12  & 1.82 & 1.90 & 1.98  & 2.14 \\
      \hline
      \multicolumn{12}{|c|}{SPMLE\_G, rare} \\
      \hline  
      Bias      & 0.02 & 0.00  & 0.00 & 0.01 & -0.01 & 0.02 & -0.02 & 0.00 & 0.00 & -0.01 & 0.01 \\
      CI(\%)    & 94.9 & 95.7  & 94.9 & 95.9 & 96.3  & 94.8 & 94.1  & 95.5 & 94.3 & 95.0  & 95.7 \\
      MSE Eff   & 1.35 & 1.29  & 1.31 & 1.35 & 1.35  & 1.27 & 2.25  & 1.94 & 2.04 & 2.10  & 2.32 \\
      \hline
      \multicolumn{12}{|c|}{Composite Likelihood Estimator, known $\pi_1$} \\
      \hline 
      Bias      & 0.01 & 0.00  & 0.00 & 0.00 & -0.01 & 0.00 & 0.00  & 0.00 & 0.00 & 0.00  & 0.01 \\
      CI(\%)    & 94.9 & 95.8  & 94.9 & 96.1 & 96.5  & 95.4 & 94.7  & 95.7 & 94.4 & 95.7  & 95.5 \\
      MSE Eff   & 1.34 & 1.26  & 1.28 & 1.33 & 1.32  & 1.28 & 2.11  & 1.81 & 1.90 & 1.98  & 2.14 \\
      \hline
      \multicolumn{12}{|c|}{Composite Likelihood Estimator, rare} \\
      \hline
      Bias      & 0.02 & 0.00  & 0.00 & 0.01 & -0.01 & 0.02 & -0.02 & 0.00 & 0.00 & -0.01 & 0.01 \\
      CI(\%)    & 94.9 & 95.6  & 94.9 & 95.9 & 96.4  & 95.2 & 94.1  & 95.8 & 94.8 & 95.3  & 95.7 \\
      MSE Eff   & 1.32 & 1.27  & 1.29 & 1.33 & 1.32  & 1.27 & 2.23  & 1.92 & 2.03 & 2.09  & 2.31 \\
      \hline
      \multicolumn{12}{|c|}{Symmetric Combination Estimator, known $\pi_1$} \\
      \hline
      Bias      & 0.00 & -0.03 & 0.00 & 0.00 & -0.01 & 0.01 & -0.03 & 0.02 & 0.00 & -0.02 & 0.01 \\
      CI$^*$(\%)& 96.7 & 95.7  & 96.7 & 96.5 & 97.8  & 95.4 & 94.8  & 96.7 & 96.2 & 96.6  & 97.2 \\
      MSE Eff   & 1.92 & 1.71  & 2.00 & 1.83 & 2.05  & 1.31 & 2.84  & 2.51 & 2.99 & 2.68  & 3.34 \\
      \hline
      \multicolumn{12}{|c|}{Symmetric Combination Estimator, rare} \\
      \hline
      Bias      & 0.01 & -0.02 & 0.00 & 0.01 & -0.01 & 0.02 & -0.05 & 0.02 & 0.00 & -0.03 & 0.00 \\
      CI$^*$(\%)& 96.4 & 95.7  & 95.7 & 96.3 & 98.1  & 94.9 & 94.0  & 97.0 & 96.4 & 95.5  & 97.5 \\
      MSE Eff   & 1.86 & 1.71  & 1.92 & 1.78 & 1.95  & 1.27 & 2.75  & 2.66 & 3.08 & 2.69  & 3.58 \\
      \hline
  \end{tabular}
  \floatfoot{\baselineskip=11pt \emph{CI}: coverage of a 95\% nominal confidence interval, calculated using asymptotic standard error.  
                                \emph{CI$^*$}: coverage of a 95\% nominal confidence interval, calculated using 200 bootstrap samples.  
                                \emph{MSE Eff}: mean squared error efficiency when compared to logistic regression.}}
\end{table}

\subsection{Simulation when the disease rate is misspecified}\label{sec:Misspec}

\Cref{tab:Misspec} presents the results of a simulation to evaluate the robustness of our method to misspecification of the population disease rate.  A sample of 1000 cases and 1000 controls was simulated using the same scenario as described in \cref{sec:SimScenario} except the logistic intercept was modified to yield true population disease rates of 0.05, 0.085, and 0.12.  In each instance, 1000 data sets were simulated and the Symmetric Combination Estimator was calculated with misspecified ``known $\pi_1=0.03$'' and again using the rare disease approximation.

When using the rare disease approximation, coverage remains near nominal until the true disease rate reached 0.085, and even then the lowest coverage rate was 91.3\% (for interaction parameter $\beta_{XG1}$, which still demonstrated a mean squared error efficiency of 2.51 compared to logistic regression).  When the disease rate was assumed ``known $\pi_1=0.03$'', nominal coverage was seen except when the population disease rate was 0.12.  This indicates the Symmetric Combination Estimator is fairly robust to disease rate misspecification, and even an imprecise estimate of $\pi_1$ is likely to be sufficient to conduct a valid analysis.
\begin{table}[!ht]
\ttabbox
  {\caption{\baselineskip=12pt Results of simulations as described in \cref{sec:SimScenario}, but with population disease rates (0.05, 0.085, 0.12).  For each disease rate, we simulated 1000 data sets and compared logistic regression, our method with misspecified ``known $\pi_1=0.03$'', and our method using the rare disease approximation.} \label{tab:Misspec}}
  {\begin{tabular}{ | l | l l l l l | l | l l l l l | }
      \hline
      &$\beta_{G1}$ &$\beta_{G2}$&$\beta_{G3}$ & $\beta_{G4}$ & $\beta_{G5}$ & $\beta_{X}$ & $\beta_{XG1}$ & $\beta_{XG2}$ & $\beta_{XG3}$ &$\beta_{XG4}$ & $\beta_{XG5}$ \\
      \hline
      True      & 0.18 & 0.18  & 0.00 & 0.18 & 0.00  & 0.41 & 0.26  & 0.00 & 0.00 & 0.26  & 0.00 \\
      \hline
      \multicolumn{4}{l}{\bf Disease Rate = 0.05}&\multicolumn{8}{l}{Logistic Regression} \\
      \hline
      Bias      & 0.00 & 0.00  & 0.00 & 0.00 & -0.01 & 0.00 & 0.00  & 0.00 & 0.00 & 0.00  & 0.01 \\
      CI(\%)    & 95.8 & 95.2  & 95.9 & 94.7 & 94.4  & 95.6 & 95.7  & 95.5 & 95.3 & 94.8  & 95.3 \\
      \hline
      \multicolumn{4}{|l}{}&\multicolumn{8}{l|}{Symmetric Combination Estimator, ``known $\pi_1=0.03$''} \\
      \hline
      Bias      & 0.00 & -0.03 & 0.00 & 0.00 & -0.01 & 0.03 & -0.05 & 0.02 & 0.00 & -0.04 & 0.01 \\
      CI$^*$(\%)& 97.6 & 94.1  & 97.0 & 94.8 & 95.8  & 95.1 & 93.7  & 95.6 & 96.8 & 94.8  & 96.8 \\
      MSE Eff   & 1.84 & 1.74  & 2.07 & 1.69 & 1.97  & 1.30 & 2.61  & 2.65 & 3.14 & 2.37  & 2.97 \\
      \hline
      \multicolumn{4}{|l}{}&\multicolumn{8}{l|}{Symmetric Combination Estimator, rare} \\
      \hline
      Bias      & 0.01 & -0.02 & 0.00 & 0.01 & -0.01 & 0.04 & -0.06 & 0.02 & 0.00 & -0.05 & 0.00 \\
      CI$^*$(\%)& 96.9 & 94.3  & 97.4 & 94.4 & 96.1  & 94.7 & 92.2  & 95.4 & 96.7 & 93.4  & 96.6 \\
      MSE Eff   & 1.75 & 1.73  & 2.03 & 1.60 & 1.89  & 1.22 & 2.48  & 2.76 & 3.22 & 2.25  & 3.11 \\
      \hline
      \multicolumn{4}{l}{\bf Disease Rate = 0.085}&\multicolumn{8}{l}{Logistic Regression} \\
      \hline
      Bias      & -0.01& 0.01  & 0.00 & 0.00 & 0.00  & 0.00 & 0.01  & -0.01 & 0.00 & 0.00  & 0.01 \\
      CI(\%)    & 94.3 & 94.9  & 95.4 & 94.4 & 93.5  & 94.5 & 94.8  & 94.0  & 95.1 & 95.8  & 94.5 \\
      \hline
      \multicolumn{4}{|l}{}&\multicolumn{8}{l|}{Symmetric Combination Estimator, ``known $\pi_1=0.03$''} \\
      \hline
      Bias      & 0.00 & -0.02 & 0.00 & 0.01 & -0.01 & 0.05 & -0.07 & 0.01 & 0.00 & -0.06 & 0.00 \\
      CI$^*$(\%)& 96.4 & 95.2  & 96.9 & 95.7 & 95.7  & 93.6 & 92.7  & 96.0 & 97.6 & 92.9  & 97.1 \\
      MSE Eff   & 1.84 & 1.81  & 1.99 & 1.61 & 1.90  & 1.18 & 2.65  & 2.81 & 3.18 & 2.15  & 3.20 \\
      \hline
      \multicolumn{4}{|l}{}&\multicolumn{8}{l|}{Symmetric Combination Estimator, rare} \\
      \hline
      Bias      & 0.01 & -0.01 & 0.00 & 0.02 & -0.01 & 0.06 & -0.08 & 0.01 & 0.00 & -0.06 & 0.00 \\
      CI$^*$(\%)& 96.5 & 95.8  & 96.5 & 95.7 & 95.3  & 92.5 & 91.3  & 96.1 & 97.3 & 91.8  & 97.1 \\
      MSE Eff   & 1.77 & 1.81  & 1.98 & 1.59 & 1.86  & 1.12 & 2.51  & 2.91 & 3.21 & 2.07  & 3.30 \\
      \hline
      \multicolumn{4}{l}{\bf Disease Rate = 0.12}&\multicolumn{8}{l}{Logistic Regression} \\
      \hline
      Bias      & 0.00 & 0.01  & -0.01& 0.00 & 0.00  & 0.00 & 0.00  & 0.00 & 0.00 & 0.00  & 0.00 \\
      CI(\%)    & 94.6 & 95.4  & 94.9 & 94.8 & 93.7  & 95.7 & 94.4  & 95.9 & 94.8 & 94.8  & 94.8 \\
      \hline
      \multicolumn{4}{|l}{}&\multicolumn{8}{l|}{Symmetric Combination Estimator, ``known $\pi_1=0.03$''} \\
      \hline
      Bias      & 0.00 & -0.02 & 0.00 & 0.02 & -0.01 & 0.06 & -0.08 & 0.01 & 0.00 & -0.07 & 0.01 \\
      CI$^*$(\%)& 96.4 & 95.6  & 96.1 & 94.5 & 95.6  & 93.5 & 89.2  & 96.4 & 97.2 & 89.0  & 96.9 \\
      MSE Eff   & 1.83 & 1.71  & 1.95 & 1.59 & 1.86  & 1.08 & 2.33  & 2.80 & 3.07 & 1.90  & 3.02 \\
      \hline
      \multicolumn{4}{|l}{}&\multicolumn{8}{l|}{Symmetric Combination Estimator, rare} \\
      \hline
      Bias      & 0.02 & -0.01 & 0.00 & 0.03 & -0.01 & 0.07 & -0.10 & 0.01 & 0.00 & -0.08 & 0.00 \\
      CI$^*$(\%)& 95.6 & 96.1  & 96.4 & 94.5 & 95.1  & 91.8 & 86.0  & 96.6 & 97.0 & 87.4  & 96.0 \\
      MSE Eff   & 1.72 & 1.72  & 1.91 & 1.53 & 1.79  & 0.99 & 2.11  & 2.95 & 3.14 & 1.78  & 3.11 \\
      \hline
  \end{tabular}
  \floatfoot{\baselineskip=11pt \emph{CI}: coverage of a 95\% nominal confidence interval, calculated using asymptotic standard error.  
                                \emph{CI$^*$}: coverage of a 95\% nominal confidence interval, calculated using 200 bootstrap samples. \\ 
                                \emph{MSE Eff}: mean squared error efficiency when compared to logistic regression.}}
\end{table}

\clearpage

\subsection{Violations of the Gene-Environment Independence Assumption} \label{sec:ViolAssump}

\Cref{tab:ViolAssump} presents the results of simulations to examine the robustness of our methods to violations of the gene-environment independence assumption.  In these simulations, a sample of 1000 cases and 1000 controls is simulated with genetic variables as described in \cref{sec:SimScenario}, but the environmental variable is normally distributed with mean $\alpha G_1$, $\alpha G_2$, or $\alpha G_3$.  We set $\alpha=0.032$ to induce dependence between $X$ and $G_j$ with $R^2=0.001$.  Here $\beta_G = \{\log(1.2), \log(1.2), 0, \log(1.2), 0\}$ as in \cref{sec:SimScenario}, but $\beta_X = \log(1.35)$, and $\beta_{GX}=\{\log(1.21), 0, 0, \log(1.21), 0\}$.  In each simulation, the logistic intercept was selected to give a population disease rate of 0.03.  In the first simulation, $X$ is correlated with $G_1$, which has a nonzero main effect and a nonzero interaction; in the second simulation, $X$ is correlated with $G_2$, which has a nonzero main effect but no interaction effect; in the third simulation, $X$ is correlated with $G_3$, which has neither main nor interaction effects.

We find that violating the gene-environment independence assumption induces bias in the estimate of the interaction parameter of the environmental variable and the specific SNP that is in violation of the gene-environment independence assumption, while the estimated interaction parameters of the other SNPs are unaffected.  When $\pi_1$ is known, estimates of the main effects of the SNP that is in violation of the gene-environment independence assumption are uncompromised.

\begin{table}[!ht]
\ttabbox
  {\caption{\baselineskip=12pt Results of simulations violating the gene-environment independence assumption with $X \sim N(0,\; 0.032G_j)$ for SNPs ($G_1$, $G_2$, $G_3$).  In each instance, we simulated 1000 data sets and compared our method, both with known $\pi_1$ and using the rare disease approximation, to logistic regression.} \label{tab:ViolAssump}}
  {\begin{tabular}{ | l | l l l l l | l | l l l l l | }
      \hline
      &$\beta_{G1}$ &$\beta_{G2}$&$\beta_{G3}$ & $\beta_{G4}$ & $\beta_{G5}$ & $\beta_{X}$ & $\beta_{XG1}$ & $\beta_{XG2}$ & $\beta_{XG3}$ &$\beta_{XG4}$ & $\beta_{XG5}$ \\
      \hline
      True      & 0.18 & 0.18  & 0.00 & 0.18 & 0.00 & 0.30  & 0.19 & 0.00 & 0.00 & 0.19  & 0.00 \\
      \hline
      \multicolumn{4}{l}{\bf $X$ correlated with $G_1$}&\multicolumn{8}{l}{Logistic Regression} \\
      \hline
      Bias      & 0.00 & 0.00  & 0.00 & 0.00 & 0.00 & 0.00  & 0.01 & 0.00 & 0.00 & 0.00  & 0.01 \\
      CI(\%)    & 95.3 & 94.9  & 95.3 & 94.3 & 93.6 & 95.4  & 95.5 & 94.0 & 94.9 & 94.1  & 95.6 \\
      \hline
      \multicolumn{4}{|l}{}&\multicolumn{8}{l|}{Symmetric Combination Estimator, known $\pi_1$} \\
      \hline
      Bias      & 0.00 & -0.03 & 0.00 & 0.00 & 0.00 & -0.01 & 0.05 & 0.01 & 0.00 & -0.02 & 0.00 \\
      CI$^*$(\%)& 95.8 & 93.3  & 95.6 & 94.3 & 95.4 & 94.2  & 92.5 & 92.7 & 95.6 & 93.5  & 95.4 \\
      MSE Eff   & 1.31 & 1.13  & 1.37 & 1.26 & 1.44 & 1.34  & 1.91 & 2.40 & 2.71 & 2.41  & 2.91 \\
      \hline
      \multicolumn{4}{|l}{}&\multicolumn{8}{l|}{Symmetric Combination Estimator, rare} \\
      \hline
      Bias      & 0.00  & -0.03 & 0.00 & 0.00 & 0.00  & 0.01  & 0.01  & 0.01 & 0.00 & -0.03 & 0.00 \\
      CI$^*$(\%)& 96.4  & 92.5  & 96.3 & 94.8 & 95.7  & 95.5  & 95.3  & 93.5 & 95.8 & 90.3  & 95.3 \\
      MSE Eff   & 1.35  & 1.14  & 1.47 & 1.34 & 1.51  & 1.43  & 3.24  & 2.67 & 2.96 & 2.27  & 3.59 \\
      \hline
      \multicolumn{4}{l}{\bf $X$ correlated with $G_2$}&\multicolumn{8}{l}{Logistic Regression} \\
      \hline
      Bias      & 0.00  & 0.00  & 0.00 & 0.00 & 0.00 & 0.00  & 0.00  & 0.00 & 0.00 & 0.01  & 0.00 \\
      CI(\%)    & 94.8  & 95.1  & 94.5 & 94.5 & 95.3 & 96.2  & 94.3  & 93.4 & 94.7 & 95.3  & 95.2 \\
      \hline
      \multicolumn{4}{|l}{}&\multicolumn{8}{l|}{Symmetric Combination Estimator, known $\pi_1$} \\
      \hline
      Bias      & -0.01 & -0.02 & 0.00 & 0.00 & 0.00 & -0.03 & -0.02 & 0.06 & 0.00 & -0.02 & 0.00 \\
      CI$^*$(\%)& 95.6  & 95.2  & 95.8 & 95.5 & 96.9 & 93.1  & 94.2  & 83.5 & 95.3 & 94.1  & 96.1 \\
      MSE Eff   & 1.33  & 1.27  & 1.35 & 1.32 & 1.42 & 1.11  & 2.79  & 1.54 & 2.84 & 2.63  & 3.21 \\
      \hline
      \multicolumn{4}{|l}{}&\multicolumn{8}{l|}{Symmetric Combination Estimator, rare} \\
      \hline
      Bias      & -0.01 & -0.02 & 0.00 & 0.00 & 0.00  & -0.01 & -0.05 & 0.05 & 0.00 & -0.03 & 0.00 \\
      CI$^*$(\%)& 95.8  & 94.9  & 95.9 & 95.1 & 96.3  & 96.0  & 87.7  & 84.7 & 95.3 & 90.3  & 97.0 \\
      MSE Eff   & 1.34  & 1.27  & 1.44 & 1.35 & 1.45  & 1.37  & 2.41  & 1.79 & 3.21 & 2.43  & 3.95 \\
      \hline
      \multicolumn{4}{l}{\bf $X$ correlated with $G_3$}&\multicolumn{8}{l}{Logistic Regression} \\
      \hline
      Bias      & 0.00  & 0.00  & 0.00 & 0.00 & -0.01 & 0.00  & 0.01  & 0.00 & 0.00 & 0.00  & 0.01 \\
      CI(\%)    & 94.9  & 94.8  & 96.0 & 94.9 & 95.2  & 95.0  & 95.9  & 95.0 & 95.6 & 94.7  & 94.3 \\
      \hline
      \multicolumn{4}{|l}{}&\multicolumn{8}{l|}{Symmetric Combination Estimator, known $\pi_1$} \\
      \hline
      Bias      & -0.01 & -0.02 & 0.01 & 0.00 & -0.01 & -0.03 & -0.01 & 0.01 & 0.05 & -0.02 & 0.00 \\
      CI$^*$(\%)& 96.0  & 93.8  & 96.3 & 96.4 & 96.5  & 92.2  & 93.5  & 95.6 & 89.9 & 94.0  & 94.8 \\
      MSE Eff   & 1.33  & 1.16  & 1.34 & 1.25 & 1.34  & 1.15  & 2.56  & 2.62 & 1.63 & 2.33  & 2.89 \\
      \hline
      \multicolumn{4}{|l}{}&\multicolumn{8}{l|}{Symmetric Combination Estimator, rare} \\
      \hline
      Bias      & -0.01 & -0.03 & 0.01 & 0.00 & -0.01 & -0.01 & -0.04 & 0.01 & 0.04 & -0.03 & 0.00 \\
      CI$^*$(\%)& 95.6  & 93.5  & 96.3 & 96.3 & 96.5  & 94.4  & 88.2  & 96.5 & 89.1 & 90.8  & 95.5 \\
      MSE Eff   & 1.41  & 1.16  & 1.35 & 1.28 & 1.37  & 1.39  & 2.37  & 3.01 & 1.78 & 2.17  & 3.62 \\
      \hline
  \end{tabular}
  \floatfoot{\baselineskip=11pt \emph{CI}: coverage of a 95\% nominal confidence interval, calculated using asymptotic standard error.  
                                \emph{CI$^*$}: coverage of a 95\% nominal confidence interval, calculated using 100 bootstrap samples. \\ 
                                \emph{MSE Eff}: mean squared error efficiency when compared to logistic regression.}}
\end{table}

\clearpage

\subsection{Simulations with alternative distributions for $G$ and $X$} \label{sec:AltDist}

\Cref{tab:AltDist} presents the results of a simulation in which $X$ and $G$ are both multivariate with a combination of discrete and continuous components.  $G_1$ and $G_2$ are correlated SNPs in Hardy-Weinberg equilibrium with minor allele frequencies (0.2, 0.3), and $G_3$ has a gamma distribution with shape = 20 and scale = 20 (to simulate a skewed polygenic risk score).  $X_1$ is binary with frequency 0.5 and $X_2$ has a standard normal distribution.  Here $\beta_G = \{\log(1.2), 0, \log(1.38)\}$, $\beta_X = \{\log(1.5), \log(1.14)\}$, $\beta_{GX}=\{\log(1.1), 0, 0, 0, 0, 0\}$, and the logistic intercept was selected to give a population disease rate of 0.05.  Using these settings, 1000 data sets were simulated with 1000 cases and 1000 controls each.

\begin{table}[!ht]
\ttabbox
  {\caption{\baselineskip=12pt Results of 1000 simulations with multivariate $G$ and $X$, comparing the bias, coverage, and efficiency of standard logistic regression to our Symmetric Combination Estimator, both with known $\pi_1$ and using the rare disease approximation.} \label{tab:AltDist}}
  {\begin{tabular}{ | l | l l l l l l l l l l l | }
      \hline
      &$\beta_{G1}$   &$\beta_{G2}$   &$\beta_{G3}$   &$\beta_{X1}$   &$\beta_{X2}$   &$\beta_{X1G1}$ &$\beta_{X1G2}$ &$\beta_{X1G3}$ &$\beta_{X2G1}$ &$\beta_{X2G2}$ &$\beta_{X2G3}$ \\
      \hline
      True      & 0.18  & 0.00  & 0.32  & 0.41  & 0.14  & 0.10  & 0.00  & 0.00  & 0.00  & 0.00  & 0.00  \\
      \hline
      \multicolumn{12}{|c|}{Logistic Regression} \\
      \hline
      Bias      & 0.01  & -0.01 & 0.00  & 0.00  & 0.00  & -0.01 & 0.00  & 0.00  & 0.00  & 0.00  & 0.00  \\
      CI(\%)    & 94.4  & 94.3  & 95.2  & 94.6  & 94.0  & 95.7  & 94.7  & 94.7  & 94.5  & 95.4  & 94.6  \\
      \hline
      \multicolumn{12}{|c|}{Symmetric Combination Estimator, known $\pi_1$} \\
      \hline 
      Bias      & -0.02 & 0.00  & -0.06 & -0.01 & -0.02 & -0.02 & 0.00  & 0.01  & 0.00  & 0.00  & 0.00  \\
      CI$^*$(\%)& 93.4  & 94.7  & 94.9  & 94.8  & 96.0  & 94.1  & 95.4  & 96.1  & 95.8  & 96.9  & 96.5  \\
      MSE Eff   & 1.48  & 1.66  & 1.61  & 2.44  & 2.75  & 2.22  & 2.67  & 2.84  & 2.90  & 2.80  & 3.03  \\
      \hline
      \multicolumn{12}{|c|}{Symmetric Combination Estimator, rare} \\
      \hline
      Bias      & -0.01 & 0.00  & -0.06 & 0.00  & -0.02 & -0.03 & 0.00  & 0.01  & 0.00  & 0.00  & 0.00  \\
      CI$^*$(\%)& 93.3  & 94.5  & 95.3  & 94.8  & 95.8  & 94.1  & 95.6  & 96.3  & 95.6  & 96.9  & 96.2  \\
      MSE Eff   & 1.54  & 1.73  & 1.66  & 2.60  & 2.98  & 2.37  & 2.98  & 3.08  & 3.25  & 3.07  & 3.32  \\
      \hline
  \end{tabular}
  \floatfoot{\baselineskip=11pt \emph{CI}: coverage of a 95\% nominal confidence interval, calculated using asymptotic standard error. \\
                                \emph{CI$^*$}: coverage of a 95\% nominal confidence interval, calculated using 100 bootstrap samples. \\ 
                                \emph{MSE Eff}: mean squared error efficiency when compared to logistic regression.}}
\end{table}

\subsection{Creating the polygenic risk score for the PLCO data analysis} \label{sec:PRSweights}

\Cref{tab:PRSweights} displays the SNPs used in the calculation of the polygenic risk score for the analysis of the Prostate, Lung, Colorectal and Ovarian cancer screening trial data described in \cref{sec:DataData}.

\begin{table}[ht]
\begin{center}
\caption{\baselineskip=12pt SNPs involved in creating the polygenic risk score, and their regression coefficients}	\label{tab:PRSweights}
\begin{tabular}{lr}
{\bf RS Number} & {\bf Coefficient} \\
    rs11249433  &    -0.02813492    \\
    rs1045485   &    -0.09307971    \\
    rs13387042	&    -0.26203658    \\
    rs4973768	&     0.08013260    \\
    rs10069690	&     0.06459363    \\
    rs10941679	&     0.09185539    \\
    rs889312	&    -0.00565121    \\
    rs17530068	&     0.09668742    \\
    rs2046210	&     0.09851217    \\
    rs1562430	&    -0.14871719    \\
    rs1011970	&     0.05329783    \\
    rs865686	&    -0.02913340    \\
    rs2380205	&    -0.01821032    \\
    rs10995190	&    -0.04275836    \\
    rs2981582	&     0.14008397    \\
    rs909116	&     0.04955235    \\
    rs614367	&     0.06438418    \\
    rs3803662	&     0.27080105    \\
    rs6504950	&    -0.17586244    \\
    rs8170      &     0.08570773    \\
    rs999737    &    -0.13737833    \\
\end{tabular}
\end{center}
\end{table}

\end{document}